\documentclass[submission,copyright,creativecommons]{eptcs}
\providecommand{\event}{CREST 2018} 

\title{Interactions between Causal Structures in \\
Graph Rewriting Systems}

\author{Ioana Cristescu
  \qquad Walter Fontana
  \institute{Department of Systems Biology, Harvard Medical School, Boston, USA}
  \email{\{ioana\_cristescu,walter\_fontana\}@hms.harvard.edu}
  \and
  Jean Krivine
  \institute{IRIF, Universite Paris 7, Paris, France}
\email{\quad jean.krivine@irif.fr}
}
\def\titlerunning{Interactions between Causal Structures}
\def\authorrunning{I. Cristescu, W. Fontana \& J. Krivine}

\usepackage{stmaryrd,amsfonts,latexsym,url,amsmath,amssymb,amscd,mathtools}
\usepackage{amsthm}
\usepackage{float}
\usepackage{graphicx}
\usepackage{tikz}
\usetikzlibrary{matrix,arrows,decorations.pathmorphing}
\usetikzlibrary{shapes}
\usetikzlibrary{calc}
\usepackage{wrapfig}
\usepackage{hyperref}
\usepackage[font={scriptsize}]{subcaption}
\usepackage[font={scriptsize}]{caption}

\newcommand{\labl}{\ensuremath{\ell}}
\def\spo{\ensuremath{\mathsf{mix}}}
\newcommand{\multisum}{\ensuremath{\mathsf{multisum}}}

\newcommand{\encc}[1]{\ensuremath{[\![#1]\!]}}
\newcommand{\enc}[1]{\encc{#1}} 
\newcommand{\enct}[1]{\ensuremath{\{#1\}}}

\newcommand{\absa}{\ensuremath{\mathcal{A}_1}}
\newcommand{\absb}{\ensuremath{\mathcal{A}_2}}
\newcommand{\concra}{\ensuremath{\mathcal{C}_1}}
\newcommand{\concrb}{\ensuremath{\mathcal{C}_2}}
\newcommand{\concre}{\ensuremath{\mathsf{concrete}}}
\newcommand{\concr}{\ensuremath{\mathcal{C}}}

\newcommand{\emb}{\ensuremath{\rightarrow}}
\newcommand{\remb}{\ensuremath{\leftarrow}}
\newcommand{\lemb}{\ensuremath{\rightarrow}}
\newcommand{\pmorph}{\ensuremath{\rightharpoonup}}
\newcommand{\vectorf}[1]{\ensuremath{\overset{\rightarrow}{#1}}}
\newcommand{\vectorm}[2]{\ensuremath{\langle #1,#2\rangle}}

\newcommand{\iso}{\ensuremath{\cong}}
\def\posI#1{\redl{+\vectorf{#1}}}
\def\negI#1{\redl{-\vectorf{#1}}}

\newcommand{\kn}{\ensuremath{\mathsf{K}}}
\newcommand{\agset}{\ensuremath{\mathsf{A}}}
\newcommand{\type}{\ensuremath{\mathsf{type}}}
\newcommand{\site}{\ensuremath{\mathsf{site}}}
\newcommand{\free}{\ensuremath{\mathrm{free}}}
\newcommand{\aga}{\ensuremath{\mathrm{a}}}
\newcommand{\agb}{\ensuremath{\mathrm{b}}}

\newcommand{\ag}{\ensuremath{\mathcal{A}}}

\newcommand{\links}{\ensuremath{\mathcal{E}}}
\newcommand{\nodes}{\ensuremath{\mathcal{N}}}

\newcommand{\set}[1]{\ensuremath{\{#1\}}}
\newcommand{\redld}[1]{\mathrel{\cdot \xrightarrow{#1} \cdot}}
\newcommand{\redl}[1]{\mathrel{\xrightarrow{#1}}}
\newcommand{\nat}{\ensuremath{\mathbb{N}}}
\newcommand{\homg}{\ensuremath{\text{hom}(\mathcal{G})}}
\newcommand{\spang}{\ensuremath{\text{span}(\mathcal{G})}}
\newcommand{\no}[1]{\ensuremath{\mathtt{#1}}}

\newcommand{\arrpos}[1]{\ensuremath{\overset{#1}{\leadsto}}}

\theoremstyle{definition}
\newtheorem{definition}{Definition}
\newtheorem{example}{Example}
\newtheorem{lemma}{Lemma}
\newtheorem{theorem}{Theorem}

\begin{document}

\maketitle

\begin{abstract}
Graph rewrite formalisms are a powerful approach to modeling complex molecular systems. They capture the intrinsic concurrency of molecular interactions, thereby enabling a formal notion of mechanism (a partially ordered set of events) that explains how a system achieves a particular outcome given a set of rewrite rules. It is then useful to verify whether the mechanisms that emerge from a given model comply with empirical observations about their mutual interference. In this work, our objective is to determine whether a specific event in the mechanism for achieving X prevents or promotes the occurrence of a specific event in the mechanism for achieving Y. Such checks might also be used to hypothesize rules that would bring model mechanisms in compliance with observations. We define a rigorous framework for defining the concept of interference (positive or negative) between mechanisms induced by a system of graph-rewrite rules and for establishing whether an asserted influence can be realized given two mechanisms as an input.
\end{abstract}

\section{Introduction}

A persistent challenge across molecular biology is to understand how a multitude of diverse and asynchronous interactions between molecular entities give rise to coherent system behavior. One difficulty arises from the combinatorial complexity inherent in chemistry: A reaction (or interaction) between structured entities, such as molecules, consists in the transformation of specific parts in a manner that depends on a few rather than all aspects specifying the reactants. Combinatorial complexity then arises because a given reactant combination can exhibit several distinct reactive patterns and the same pattern can occur across many distinct reactant combinations. This idea generalizes beyond chemistry.

A molecular system can thus be described in terms of rewrite \emph{rules}. In this way, rule-based modeling tackles combinatorial complexity without succumbing to it because it only specifies rules of pattern transformation and not the multitude of possible carriers of these patterns. Many physical systems can be conveniently described as graphs. A rule-based approach then becomes a graph rewriting formalism with a domain-specific execution model that determines the probability with which a rule fires at a given time. The currently most developed approaches are the Kappa language~\cite{Danos2004,DanosFFHK08} and BNGL~\cite{bngl} for molecular biology and M{\o}d~\cite{moll} for organic chemistry.

A rule formalizes the interaction between physical entities at some chosen level of abstraction. Processes occurring below that level are abstracted away, yet not ignored: They inform what a rule should say, but they are not explicitly represented by it. For instance, in organic chemistry, a rule of reaction between molecules expresses a local reconfiguration of bonds among atoms without explicitly representing the underlying mechanism of electron pushing that engenders such reconfiguration. In molecular biology, an interaction between proteins is typically expressed by asserting the conditions for a change of protein state without representing the structural mechanisms enabling that change. In essence, a mechanism below the chosen abstraction level becomes an axiomatic rule at the abstraction level~\cite{siglog}.

Many observations of system behavior are \emph{assertions} rather than rules. For example, an assertion might claim that the activation of protein $\no{X}$ inhibits the assembly of molecular machine $\no{Z}$. It is desirable to determine whether and why an assertion holds in terms of the joint action among rules that represent a particular system. This is tantamount to providing a \emph{mechanism} that \emph{explains} a given assertion at the level of abstraction at which rules are defined.

The stochastic application of rules (a simulation) typically generates a long trace of state transitions. A mechanism is a set of transitions that were jointly necessary in producing a specified outcome. Mechanisms so-defined can be extracted from traces~\cite{danos07,danos12} and abstracted into partial orders (posets) of events\footnote{In Ref.\cite{danos07}, a partially ordered set of events that account for an outcome was dubbed a ``story", which is akin to the biological notion of a ``pathway".}.

Here we propose a formal logic to express and verify a particular kind of assertion about a model written in the Kappa language. We focus on assertions in which the occurrence of one event is claimed to interfere with another event.
Our approach takes as input two posets of events (i.e. mechanisms), which might be hypothesized or abstracted from a simulation, and provides evidence whether the two posets interfere with one another at the specified events. The key is that each poset builds up a context that is required for its terminal event. These contexts can be reconstructed and checked for mutual consistency. To lay the foundation for this approach requires setting up some formal machinery which occupies the bulk of this paper.

\begin{wrapfigure}{r}{0.5\textwidth}
\centering
\includegraphics[scale=0.2]{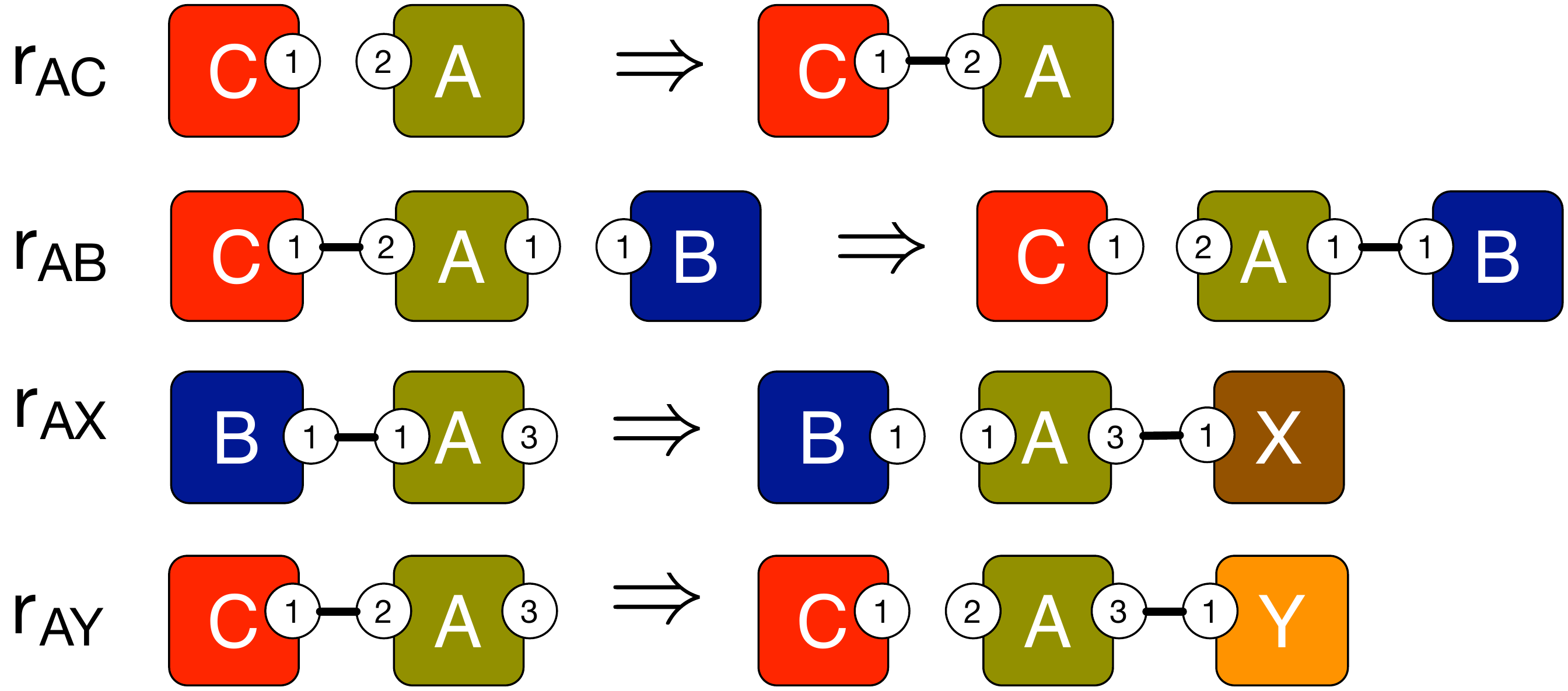}
\caption{\label{fig:model} A Kappa model.\hfill}
\end{wrapfigure}
\noindent\emph{Interaction between graph rewriting posets.}
The graphs in Kappa consist of nodes, called agents, meant to represent proteins. Agents are equipped with \emph{sites} through which they connect to one another. A site represents a resource and hence can bear at most one edge. Such graphs are called \emph{site-graphs}.

\begin{figure}
  \centering
\includegraphics[scale=0.2]{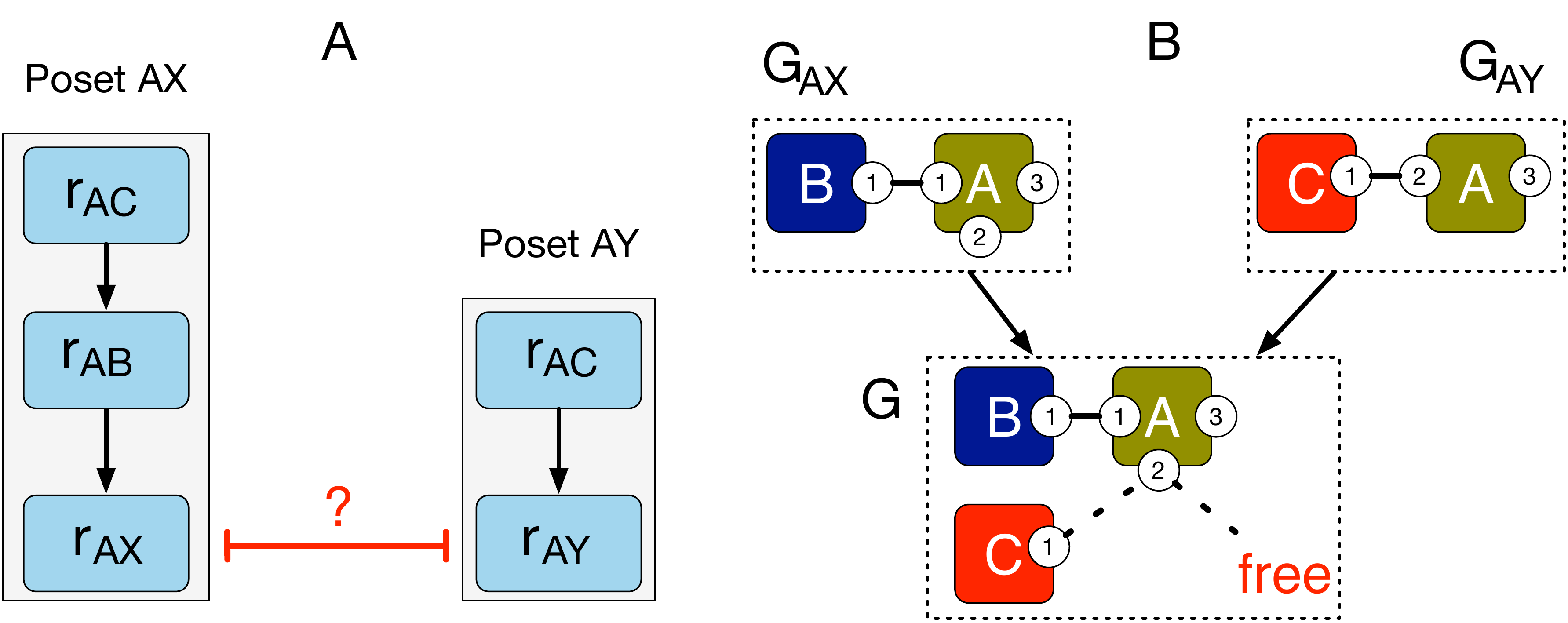}
\caption{\label{fig:conflict} Panel A: Two mechanisms (posets) and a query for conflict between the specified events. Black arrows are precedence; events are labeled by the underlying rules. Panel B: The graph $G_{\text{AX}}$ represent the context in which the rule $r_{\text{AX}}$ is applied in the Poset $\text{AX}$. There is no scenario in which the posets interact, since the graph $G$ is not a site-graph. The site $3$ of agent $\no{A}$ has to be bound to an agent $\no{C}$ and be free at the same time, which is not representable in site-graphs.}
\end{figure}

An event is the application of a rewrite rule to a usually large graph representing the state of the system. Events are partially ordered by a relation of \emph{precedence}. Intuitively, an event $e_1$ precedes an event $e_2$ if $e_1$ contributes to establishing the context necessary for $e_2$. Consider, for example, the simple model in Figure~\ref{fig:model} with the initial state consisting of nodes $\{\no{A},\no{B},\no{C}\}$, all unbound. Suppose furthermore that the binding of agent $\no{X}$ to $\no{A}$ (rule $r_\text{AX}$) and of agent $\no{Y}$ to $\no{A}$ (rule $r_\text{AY}$) are two significant events $e_{\text{AX}}$ and $e_{\text{AY}}$, respectively. We wish to verify the assertion that either event inhibits the other. The assertion is cast in terms of two mechanisms (posets) that could have been extracted from a simulation trace of this model, one mechanism resulting in $e_{\text{AX}}$, the other in $e_{\text{AY}}$ (Figure~\ref{fig:conflict}A).

\noindent A static inspection of rules $r_\text{AX}$ and $r_\text{AY}$, underlying the events that are the subject of our query, shows that both use the same site of $\no{A}$. This might suggest that the two events are in conflict and therefore inhibit each other. This, however, is not a valid conclusion. Given the poset $\text{AX}$ of Figure~\ref{fig:conflict}A, we can reconstruct the context---specifically the site-graph $G_{\text{AX}}$ of Figure~\ref{fig:conflict}B---in which rule $r_{\text{AX}}$ fires. Note that $G_{\text{AX}}$ specifies that site $2$ of $\no{A}$ must be unbound. Likewise, the firing of rule $r_{\text{AY}}$ is contingent upon context $G_{\text{AY}}$, which is built up by poset $\text{AY}$. $G_{\text{AY}}$ requires that site $2$ of $\no{A}$ be bound. These two contexts are in conflict and thus cannot be realized at the same time. This means, in turn, that there is \emph{no} inhibition \emph{at this point between the two mechanisms}: Whether a particular $\no{A}$ gets bound to $\no{X}$ or to $\no{Y}$ is already decided before the mechanisms reach the events whose relationship of inhibition we queried. As a whole, the mechanisms $\text{AX}$ and $\text{AY}$ must interfere with one another negatively, as $\no{A}$ cannot be bound to both $\no{X}$ and $\no{Y}$ at the same time; but the point of conflict is somewhere else. (It is between event $r_{\text{AB}}$ and $r_{\text{AY}}$.) To determine the earliest event combination at which two mechanisms conflict with one another can be done by scanning all events of one against all events of the other.%

\begin{wrapfigure}{R}{0.4\textwidth}
\centering
\includegraphics[scale=0.18]{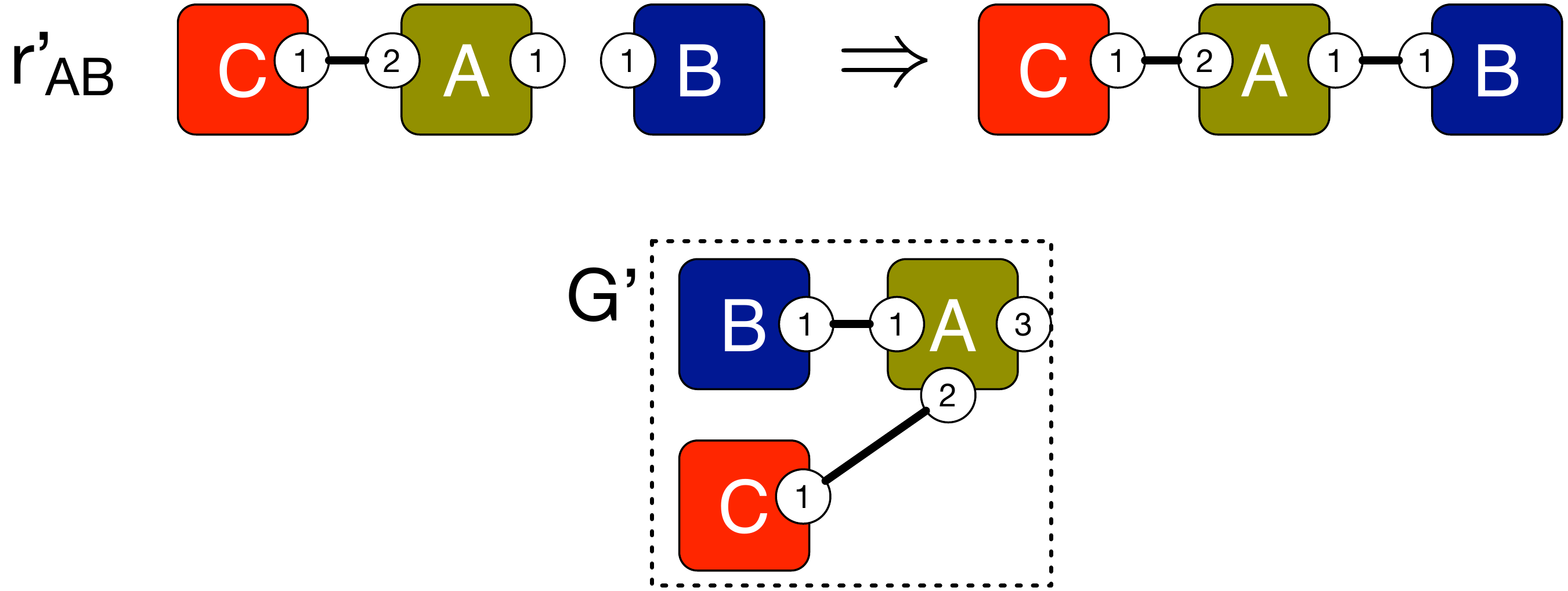}
\caption{\label{fig:inhibition} Rule $r'_{\text{AB}}$ replaces rule $r_{\text{AB}}$. The graph $G'$ represent the context in which the rule $r_{\text{AX}}$ is applied in the Poset $AX$. In this scenario, which coincides with graph $G'$, there is an inhibition between the two posets $AX$ and $AY$.}
\end{wrapfigure}
If we change the model by replacing rule $r_{\text{AB}}$ with rule $r'_{\text{AB}}$ (Figure~\ref{fig:inhibition}), the context for the application of rule $r_{\text{AX}}$ becomes consistent with the context of application of rule $r_{\text{AY}}$. This means both rules can fire. Since the firing of $r_{\text{AX}}$ destroys part of the context needed by $r_{\text{AY}}$ (and vice versa), the mechanisms inhibit each other at the events queried. In sum, the key in determining whether two events in the scope of distinct mechanisms are mutually exclusive consists in reconstructing from the given mechanisms the context required for both events and determining whether it can be realized. We call this critical context a ``scenario".

\paragraph*{Related work.} The notion of \emph{rule influence}, introduced in Refs.\@ \cite{Danos2007,a_5571737}, is used to detect inhibition and promotion between posets (Definition~\ref{def:scenario}). Our approach to abstracting traces of state transitions into partial orders is similar to Refs.~\cite{montanari93,BaldanThesis}, but we use more fine-grained relations on graphs (the enablement and prevention relations of Section~\ref{sec:ts}). As a consequence, we do not need Petri nets as an intermediate encoding between state transitions and posets. In any case, our main focus is on reconstructing a trace from a poset, which is obtained from a causal structure extracted from a Kappa simulation. This extraction is the subject of Ref.~\cite{danos12} and outside the scope of this paper.

\paragraph*{Outline.} In Section 2 we introduce site-graphs, the graph rewriting framework of Kappa, and the notion of rule influence. To define partial orders between events in a manner informed by rule influence, we need to take a detour via the transition system induced by the rules (Section 2.4). 
In order to determine the scenario that establishes enablement or prevention between posets, we need to reconstruct a trace from a poset. To this end, in Section 3, we formalize trace reconstruction as the reverse of poset abstraction from traces. In Section 4 we define a logic for expressing assertions on the posets provided as inputs. We conclude in Section 5.

Length limitations preclude a description of our implementation, which can be found at \url{https://github.com/Kappa-Dev/PosetLogic}. All constructions on site-graphs presented here can be adapted to simple graphs.

\section{Graph rewriting and transition systems}
\label{sec:ts_graph}


\subsection{Site-graphs}
\label{sec:site_graphs}

Let $\agset$ be a set of \emph{agents}, ranged over by $\aga,\agb$ and $\kn = \{A, B,..\}$ be a set of \emph{agent types}, equipped with a map $\site:\kn\to\nat_{>0}$. The function $\type:\agset\to\kn$ assigns a type to each agent.

\begin{definition}[Site-graph]
  \label{def:site_graphs}
  A site-graph is a structure $(\ag,\nodes,\links)$ where
  \begin{itemize}
  \item $\ag\subseteq \agset$ is a set of agents;
  \item $\nodes\subseteq \ag\times\nat_{>0}\uplus\{\free\}$ is a set of nodes, with a special node \free, and where each non-$\free$ node is a pair $(\aga,i)$ of an agent $\aga\in\ag$ and site $i<\site(\type(\aga))$; 
  \item $\links\subseteq \nodes\times\nodes$ is a symmetric set of edges with the constraint that it is \emph{conflict-free}: $\forall (n_1,n_2),(n_1',n_2')\in\links$, $(n_1=n_1'\wedge n_2=n_2')$ $\vee$ $(n_1=n_2'\wedge n_2=n_1')$ $\vee$ $(\{n_1,n_2\}\cap \{n_1',n_2'\} \subseteq \{\free\})$.
  \end{itemize}
\end{definition}

\begin{definition}[Morphism on site-graphs]
  \label{def:site_morph}
  A morphism $f:G\to H$, for $G$ and $H$ two site-graphs, is a pair of functions $f=(v,e)$ with
  \begin{itemize}
  \item $v:\ag_G \to \ag_H$ a function on agents that preserves types: $\type(v(\aga)) = \type(\aga)$ and that can be extended to a function on nodes: $v(\aga,i) = (v(\aga),i)$ and $v(\free)=\free$, for all $\aga\in\ag_G$ and for all $i<\site(\type(\aga))$;
  \item and $e:\links_G \to\links_H$ a function on edges such that for any two nodes $n_1,n_2\in\nodes_G$, if $(n_1,n_2)\in\links_G$ then $e(n_1,n_2)=(v(n_1),v(n_2))$.
  \end{itemize}
\end{definition}

Site-graphs and their morphisms form a category, denoted $\mathcal{G}$.
Morphisms in $\mathcal{G}$ preserve the node type and the edge structure of nodes in site-graphs.
%
Isomorphisms are denoted with $\iso$.
A \emph{mono} is a morphism with injective functions on nodes and edges.
We denote the empty graph with $\varepsilon$ and write $\vectorf{f}=\vectorm{f_1}{f_2}$ for the span $G_1\overset{f_1}{\leftarrow} G_2\overset{f_2}{\rightarrow} G_3$. The same notation is used to denote the cospan $G_1\overset{f_1}{\rightarrow} G_2\overset{f_2}{\leftarrow} G_3$.
For simplicity, we write $f$ for $v$ (or $e$) in $f=(v,e)$.
Finally, we write $\homg$ and $\spang$ for the class of morphisms and spans of $\mathcal{G}$, respectively.

\subsection{Graph rewriting}
\label{sec:dpo}
A rule-based model consists of graph-rewriting rules that are applied in a stochastic fashion to a typically large graph representing the state of a system. 
In Kappa the stochastic application of rewrite rules follows basic principles of stochastic chemical kinetics \cite{Gillespie77,Danos2007}. Each graph-rewrite action constitutes a state transition and a temporal sequence of such transitions is a trace. We also refer to the state of the system as a ``mixture".

\begin{definition}[Pushout]
The \emph{pushout} of a span $\vectorf{g}$ is a cospan $\vectorf{f}$ such that $f_1g_1=f_2g_2$\footnote{We write $fg(x) = f(g(x))$, with $x$ in the domain of $g$, for morphisms composition.} and such that for any other cospan $\vectorf{f'}$ for which $f_1'g_1=f_2'g_2$, there is a unique morphism $M\to M'$ that makes diagram PO below commute.
\end{definition}

In the category of site-graphs, the pushout does not always exist. For a span $\vectorf{g}$ of monos, if the pushout exists, then it asserts a gluing of $G_1$ and $G_2$, resulting in $M$, based on the identifications (gluing instructions) expressed by $\vectorf{g}$.

\begin{definition}[Rule]
  \label{def:rule}
  A \emph{rule} is a span of monos $\vectorf{r} = L\overset{p}{\remb} K \overset{q}{\lemb} R$ such that
  for some $\aga\in\ag_K$ and $i<\site(\type(\aga))$, if $(\aga,i)\in\nodes_K$ then $\big((q(\aga),i),n\big) \in\links_R \iff \big((p(\aga),i),n'\big) \in\links_L$, with $n\in\nodes_R$ and $n'\in\nodes_L$.
\end{definition}

In site-graphs the site of an agent can be specified without specifying if the site is free or bound to another site. Formally, in a site-graph $G$ with an agent $\aga\in\ag_G$, we can have $(\aga,i)\in\nodes_G$ for which there is no edge $(n_1,n_2)\in\links_G$ such that $n_1=(\aga,i)$ or $n_2=(\aga,i)$. Rules however, need to satisfy a constraint related to sites: if an edge exists for a site in either sides of a rule, then it exists in both sides.
\begin{definition}[Double-pushout rewriting~\cite{AlgebraicGR}]
  \label{def:dpo}
  Let $\vectorf{r} = L\overset{p}{\remb} K \overset{q}{\lemb} R$ be a rule. Let $M$ be a site-graph (typically a system state) and let $m:L\lemb M$ be a mono, called \emph{a matching}.
The \emph{double pushout rewriting} consists in defining the site-graph $D$, called the \emph{context} graph, and the site-graph $N$ such that the two squares in diagram DPO are pushouts.
We refer to the dpo rewrite of $M$ to $N$ as $M\overset{m,\vectorf{r}}{\Rightarrow}N$ and denote the state transition associated with the application of rule $\vectorf{r}=\vectorm{p}{q}$ at "location" $m$ of the system state $M$ (i.e.\@ the mixture) as $\spo(M\overset{m,\vectorf{r}}{\Rightarrow}N) = M{\remb} D {\lemb} N$.
\end{definition}

\begin{minipage}{.9\textwidth}
  \centering
    \begin{tikzpicture} [scale=0.8]
    \node (t) at (-2.5,0.5) {\small{PO:}};
    \node (mp) at (0,2) {\(M'\)};
    \node (o) at (0,-1) {\(O\)};
    \node (l1) at (-1,0) {\(G_1\)};
    \node (l2) at (1,0) {\(G_2\)};
    \node (m) at (0,1) {\(M\)};
    \draw [->] (o) -- node [below,near end] {\(g_1\)} (l1);
    \draw [->] (o) -- node [below,near end] {\(g_2\)} (l2);
    \draw [->] (l1) -- node [above,near start] {\(f_1~\)} (m);
    \draw [->] (l2) -- node [above,near start] {\(~f_2\)} (m);
    \draw [->] (l1) to [bend left] node [left] {\(f_1'\)} (mp);
    \draw [->] (l2) to [bend right] node [right] {\(f_2'\)} (mp);
    \draw [->, dotted] (m) -- (mp);
  \end{tikzpicture}
  \hspace*{1cm}
  \centering
  \begin{tikzpicture} [scale=0.8]
    \node (t) at (-3,0.7) {\small{DPO:}};
    \node (0) at (-1.5,-1) {\ };
    \node (l) at (-1.5,0) {\(L\)};
    \node (d) at (0,0) {\(K\)};
    \node (r) at (1.5,0) {\(R\)};
    \node (m) at (-1.5,1.5) {\(M\)};
    \node (d') at (0,1.5) {\(D\)};
    \node (n) at (1.5,1.5) {\(N\)};
    \draw [] (0);
    \draw [->] (d) -- node [above] {\(p\)} (l);
    \draw [->] (d) -- node [above] {\(q\)} (r);
    \draw [->] (d') -- (m);
    \draw [->] (d') -- (n);
    \draw [->] (l) -- node [left,midway] {\(m\)}  (m);
    \draw [->] (d) -- (d');
    \draw [->] (r) -- (n);
  \end{tikzpicture}
\end{minipage}

Given the definition above, a context graph $D$ need not always exist. We use dpo rewriting for the sake of simplicity, but our work extends to other graph rewriting techniques.

\subsection{Influence}
\label{sec:infl}

The postcondition resulting from the application of a rule $\vectorf{r_1}$ can satisfy or, more generally, contribute (in conjunction with other rules) to satisfying the precondition for the application of another rule $\vectorf{r_2}$.  Alternatively, $\vectorf{r_1}$ might destroy the precondition of $\vectorf{r_2}$. In the former case we speak of a positive influence of $\vectorf{r_1}$ on $\vectorf{r_2}$ and, in the latter case, of a negative influence. Of course, a rule may also have no influence on a particular other rule.

Influence\footnote{Positive and negative influence were referred to as \emph{activation}, \emph{inhibition} or \emph{overlaps} in Refs.\cite[Section 3.4]{Danos2007},\cite[Section 4.2.3]{a_5571737}\cite{RussInfluence}.} belongs to the realm of possibility: it is a latent relation between rules that becomes manifest as a relation between events (i.e.\@ actual rule applications) in the specific context of a trace, as we discuss formally in Section~\ref{sec:posets_of_graphs}.

We next define two categorical concepts needed for capturing influence. Multisums are meant to characterize all possible ways of gluing together two graphs $G_1$ and $G_2$.  

\begin{definition}[Multisum in the subcategory of monos]
  \label{def:multisum}
Let $G_1$ and $G_2$ be two graphs. The \emph{multisum} of $G_1$ and $G_2$, denoted with $\multisum(G_1,G_2)$, is a family of cospans of monos $\vectorf{f_i}=\vectorm{f_{1,i}}{f_{2,i}}$, with $f_{j,i}:G_j\lemb M_i$, $i\leq n$, $j\in\{1,2\}$,
such that for any other cospan of monos $\vectorf{f'}$, with $f_j':G_j\lemb M'$,
there exists an $M_k$, $k\leq n$ and a unique mono $M_k\lemb M'$ that makes diagram MS below commute. Moreover, for any monos $M_k\lemb M'$ and $M_i\lemb M'$, $i,k\leq n$, for which diagram MS commutes, we have $M_k\iso M_i$.
\end{definition}

Unlike other constructions, which are defined in $\mathcal{G}$, multisums are defined in the subcategory of $\mathcal{G}$ whose morphisms are restricted to monos. Multisums always exists in this subcategory.

\begin{definition}[Pullback]
The \emph{pullback} of the cospan $\vectorf{f}$ consists of a span $\vectorf{g}$ such that $f_1g_1=f_2g_2$. In addition, for any other span $\vectorf{g'}$ such that $f_1g_1'=f_2g_2'$, there is a unique morphism $O'\to O$ that makes diagram PB commute.
\end{definition}

\begin{minipage}{.9\textwidth}
\centering
    \begin{tikzpicture} [scale=0.8]
    \node (t) at (-3,1) {\small{MS:}};
    \node (0) at (-2,-0.5) {\ };
    \node (l1) at (-1.5,0) {\(G_1\)};
    \node (l2) at (1.5,0) {\(G_2\)};
    \node (m1) at (-1.5,1) {\(M_1\)};
    \node (m3) at (-0.6,1) {\(\cdots\)};
    \node (m4) at (0,1) {\(M_k\)};
    \node (m2) at (0.6,1) {\(\cdots\)};
    \node (mn) at (1.5,1) {\(M_n\)};
    \node (n) at (0,2.5) {\(M'\)};
    \draw [->] (l1) -- (m1);
    \draw [->] (l2) -- (m1);
    \draw [->] (l1) -- (mn);
    \draw [->] (l2) -- (mn);
    \draw [->] (l1) to [out=150,in=180] node [left] {\(f_1'\)} (n);
    \draw [->] (l2) to [out=30,in=0] node [right] {\(f_2'\)} (n);
    \draw [->, dotted] (m4) -- (n);
    \draw [->] (l1) -- (-0.2,0.7);
    \draw [->] (l2) -- (0.2,0.7);
  \end{tikzpicture}
\hspace*{1cm}
\centering
  \begin{tikzpicture} [scale=0.8]
    \node (t) at (-2,-0.5) {\small{PB:}};
    \node (o) at (0,-1) {\(O\)};
    \node (l1) at (-1,0) {\(G_1\)};
    \node (l2) at (1,0) {\(G_2\)};
    \node (m) at (0,1) {\(M\)};
    \node (n) at (0,-2) {\(O'\)};
    \draw [->] (o) -- node [below,near end] {\(g_1\)} (l1);
    \draw [->] (o) -- node [below,near end] {\(g_2\)} (l2);
    \draw [->] (l1) -- node [above,near start] {\(f_1~\)} (m);
    \draw [->] (l2) -- node [above,near start] {\(~f_2\)} (m);
    \draw [->] (n) to [bend left] node [left] {\(g_1'\)} (l1);
    \draw [->] (n) to [bend right] node [right] {\(g_2'\)} (l2);
    \draw [->, dotted] (n) -- (o);
  \end{tikzpicture}
\end{minipage}
\ \\

\par
The pullback always exists in $\mathcal{G}$. Using these notions we can define influence.

\begin{definition}[Positive influence~\cite{RussInfluence}]
  \label{def:pos_infl}
Given two rules $\vectorf{r_1}=L_1{\remb} K_1\overset{i}{\lemb} R_1$ and $\vectorf{r_2}=L_2{\remb} K_2 {\lemb} R_2$, consider an overlap between $R_1$ and $L_2$, i.e.\@ a cospan $\vectorf{f}\in\multisum(R_1, L_2)$, and let $\vectorf{g}$ be the pullback of $\vectorf{f}$. Moreover, let $\vectorf{h}$ be the pullback of $\vectorm{i}{g_1}$. The rule $\vectorf{r_1}$ has a \emph{positive influence} on rule $\vectorf{r_2}$, if $h_2$ is not an iso. In other words,
if $O$ is not contained in $P$ and, thus, in $K_1$.
The influence is induced by
the overlap $\vectorf{g}$ corresponding to $\vectorf{f}$ and is denoted by $\vectorf{r_1}\posI{g} \vectorf{r_2}$.
\end{definition}

\begin{wrapfigure}{R}{0.45\textwidth}
\centering
\vspace*{-20pt}
    \begin{tikzpicture} [scale=0.8]
    \node (p) at (0,-1) {\(P\)};
    \node (d1) at (-1,0) {\(K_1\)};
    \node (o) at (1,0) {\(O\)};
    \node (m) at (1,2) {\(M\)};
    \node (r1) at (0,1) {\(R_1\)};
    \node (l1) at (-2,1) {\(L_1\)};
    \node (l2) at (2,1) {\(L_2\)};
    \node (d2) at (3,0) {\(K_2\)};
    \node (r2) at (4,1) {\(R_2\)};
    \draw [->,dotted] (p) -- node [below] {\(h_1\)} (d1);
    \draw [->,dotted] (p) -- node [below,near end] {\(h_2\)} (o);
    \draw [->] (d1) -- node [above,near start] {\(i\)} (r1);
    \draw [->] (d1) -- (l1);
    \draw [->] (o) -- node [below] {\(g_1\)} (r1);
    \draw [->] (o) -- node [below,near end] {\(g_2\)} (l2);
    \draw [->] (r1) -- node [above,near start] {\(f_1~\)} (m);
    \draw [->] (l2) -- node [above,near start] {\(~f_2\)} (m);
    \draw [->] (d2) -- (r2);
    \draw [->] (d2) -- (l2);
    \end{tikzpicture}
\vspace*{-10pt}
\end{wrapfigure}
The diagram on the right depicts the relationships used in Definition \ref{def:pos_infl}. The rule $\vectorf{r_1}$ has a positive influence on $\vectorf{r_2}$ if it creates a subgraph of $L_2$. By requiring $h_2$ not to be an iso, we assert that $O$ is not already present in $L_1$ and must, therefore, be produced by $\vectorf{r_1}$.
Negative influence $\vectorf{r_1}\negI{g} \vectorf{r_2}$ is defined analogously, but with $\vectorf{g}$ now the pullback of a cospan $\vectorf{f}\in\multisum(L_1, L_2)$ between the left hand sides: $\vectorf{r_1}$ has a negative influence on $\vectorf{r_2}$, if it destroys a subgraph of $L_2$.

\hspace*{-\parindent}%
\begin{minipage}{.4\textwidth}
\begin{example}
The rule $\vectorf{r_1}$ has a positive influence on $\vectorf{r_2}$, because $\vectorf{r_1}$ produces an agent $\no{B}$ needed for a subsequent application of $\vectorf{r_2}$ shown in Figure~\ref{fig:posets7}. Similarly $\vectorf{r_2}$ has a negative influence on $\vectorf{r_1}$ since it erases an agent $\no{A}$ needed by $\vectorf{r_1}$.
\end{example}
\end{minipage}
\begin{minipage}{.6\textwidth}
\begin{center}
  \includegraphics[scale=0.4]{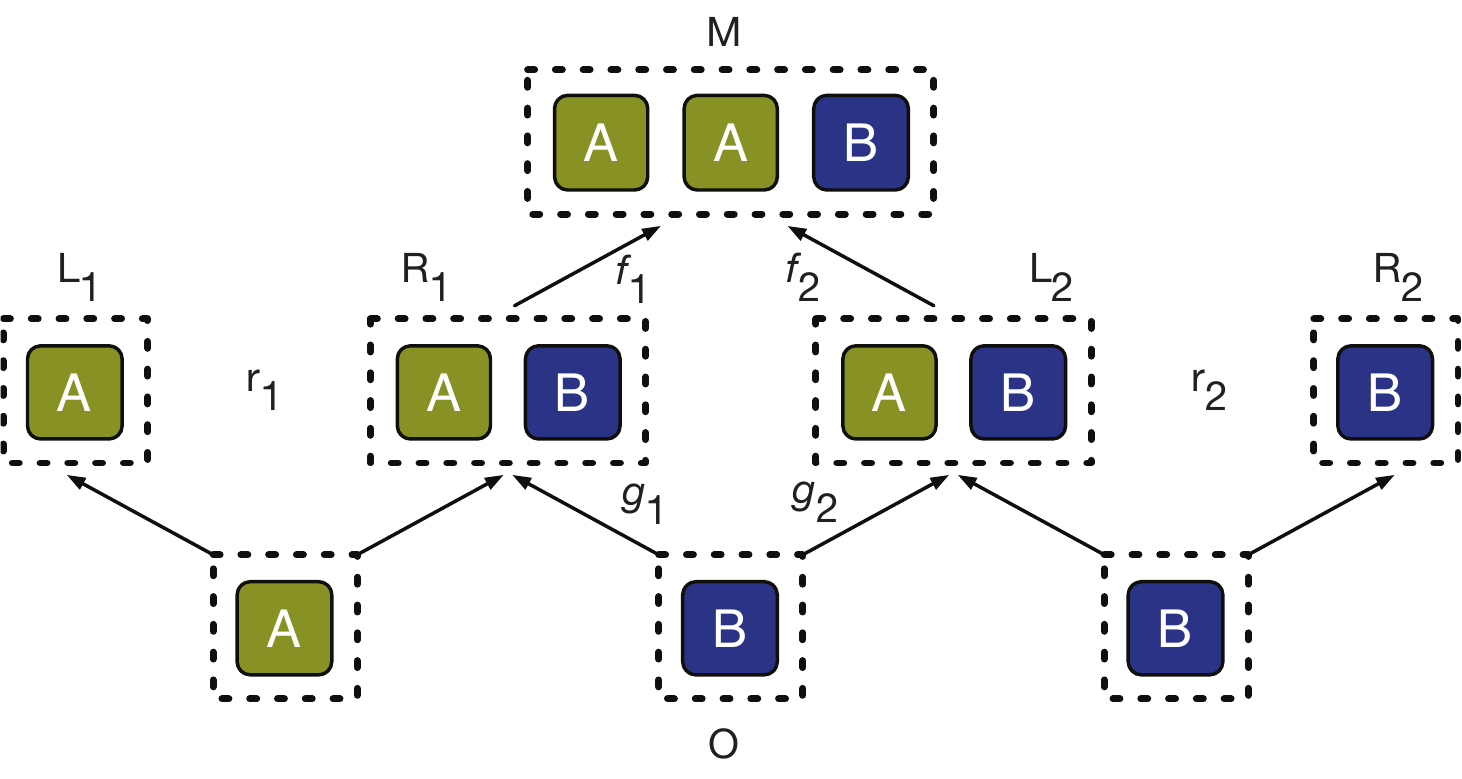}
  \captionof{figure}{Positive influence between two rules. For simplicity, sites are omitted.}
  \label{fig:posets7}
\end{center}
\end{minipage}

\subsection{Transition systems}
\label{sec:ts}

Following Refs.\@ \cite{AlgebraicGR,BaldanThesis} we introduce the notion of transition system (TS) on state graphs and an independence relation between transitions. We then propose new relations of \emph{enablement} and \emph{prevention} between transitions, based on the notions of rule influence just defined, and connect them to independence.

\begin{definition}[TS on graphs~\cite{AlgebraicGR}]
  A transition system $TS = (Q,\mathcal{R},T)$ on graphs consists of:
  \begin{itemize}
  \item a set of states $Q\subseteq \mathcal{G}$, where each state is a graph;
  \item a set of rules $\mathcal{R}$;
  \item a set of labeled transitions $T\subseteq Q\times \text{hom}(\mathcal{G})\times \mathcal{R}\times Q$, where each transition $t$ is a dpo rewriting step $M \overset{m,\vectorf{r}}{\Rightarrow} N$ with $M,\,N\in Q$, an underlying rule $\vectorf{r}:L\remb K\lemb R\in\mathcal{R}$, and a matching $m:L\emb M\in\text{hom}(\mathcal{G})$.
  \end{itemize}
\end{definition}

Transitions can be composed $t_1;t_2$ 
if the source state of $t_2$ matches the destination state of $t_1$. A \emph{trace} $\theta$ is a (possibly empty) sequence of composable transitions: $\theta=t_1;t_2;\cdots ;t_n$.

\begin{definition}[Independence relation on transitions~\cite{AlgebraicGR}]
  \label{def:indep}
  Let $t_1:M\overset{m_1,\vectorf{r_1}}{\Rightarrow} M_1$, $t_2:M_1\overset{m_2,\vectorf{r_2}}{\Rightarrow} M_2$ and $t_3:M\overset{m_3,\vectorf{r_3}}{\Rightarrow} M_3$ be transitions with underlying rules $\vectorf{r_i}=L_i{\remb} K_i {\lemb} R_i\in \mathcal{R}$, $i\in\{1,2,3\}$ and corresponding matchings $m_i$ as indicated in the diagrams below.

  \begin{description}
  \item[sequential independence]
%
$t_1$ and $t_2$ are sequentially independent, written $t_1 \Diamond_{\text{seq}} t_2$, iff there exist morphisms $i:R_1\emb D_2$ and $j:L_2\emb D_1$ such that $f_2i = n_1$ and $g_1j= m_2$. 

  \item[parallel independence]
%
$t_1$ and $t_3$ are parallel independent, written $t_1 \Diamond_{\text{par}} t_3$, iff there exist morphisms $i:L_1\emb D_3$ and $j:L_3\emb D_1$ such that $f_3i= m_1$ and $f_1j= m_3$.

    \begin{center}
    \begin{tikzpicture}[scale=0.8]
    \node (r1) at (1.5,0) {\(R_1\)};
    \node (m1) at (2,1.5) {\(M_1\)};
    \node (l2) at (2.5,0) {\(L_2\)};
    \node (d1) at (0,1.5) {\(D_1\)};
    \node (k1) at (0,0) {\(K_1\)};
    \node (d2) at (4,1.5) {\(D_2\)};
    \node (k2) at (4,0) {\(K_2\)};
    \node (l1) at (-1.5,0) {\(L_1\)};
    \node (m) at (-1.5,1.5) {\(M\)};
    \node (r2) at (5.5,0) {\(R_2\)};
    \node (m2) at (5.5,1.5) {\(M_2\)};
    \node (t) at (2,-1) {\small{sequential independence}};
    \draw [->] (k1) -- (r1);
    \draw [->] (k2) -- (l2);
    \draw [->] (k1) -- (d1);
    \draw [->] (k2) -- (d2);
    \draw [->] (d1) -- node [above,midway] {\small{\(g_1\)}} (m1);
    \draw [->] (d2) -- node [above,midway] {\small{\(f_2\)}} (m1);
    \draw [->] (l2) -- node [right,midway] {\small{\(m_2\)}} (m1);
    \draw [->] (r1) -- node [left,midway] {\small{\(n_1\)}} (m1);
    \draw [dotted,->] (r1) -- (d2);
    \draw [dotted,->] (l2) -- (d1);
    \draw [->] (k1) -- (l1);
    \draw [->] (d1) -- (m);
    \draw [->] (l1) -- node [right,midway] {\small{\(m_1\)}} (m);
    \draw [->] (k2) -- (r2);
    \draw [->] (d2) -- (m2);
    \draw [->] (r2) -- (m2);
    \end{tikzpicture}
    \hspace{0.1cm}
    \begin{tikzpicture}[scale=0.8]
    \node (r1) at (1.5,0) {\(L_1\)};
    \node (m1) at (2,1.5) {\(M\)};
    \node (l2) at (2.5,0) {\(L_3\)};
    \node (d1) at (0,1.5) {\(D_1\)};
    \node (k1) at (0,0) {\(K_1\)};
    \node (d2) at (4,1.5) {\(D_3\)};
    \node (k2) at (4,0) {\(K_3\)};
    \node (l1) at (-1.5,0) {\(R_1\)};
    \node (m) at (-1.5,1.5) {\(M_1\)};
    \node (r2) at (5.5,0) {\(R_3\)};
    \node (m2) at (5.5,1.5) {\(M_3\)};
    \node (t) at (2,-1) {\small{parallel independence}};
    \draw [->] (k1) -- (r1);
    \draw [->] (k2) -- (l2);
    \draw [->] (k1) -- (d1);
    \draw [->] (k2) -- (d2);
    \draw [->] (d1) -- node [above,midway] {\small{\(f_1\)}} (m1);
    \draw [->] (d2) -- node [above,midway] {\small{\(f_3\)}} (m1);
    \draw [->] (l2) -- node [right,midway] {\small{\(m_3\)}} (m1);
    \draw [->] (r1) -- node [left,midway] {\small{\(m_1\)}} (m1);
    \draw [dotted,->] (r1) -- (d2);
    \draw [dotted,->] (l2) -- (d1);
    \draw [->] (k1) -- (l1);
    \draw [->] (d1) -- (m);
    \draw [->] (l1) -- (m);
    \draw [->] (k2) -- (r2);
    \draw [->] (d2) -- (m2);
    \draw [->] (r2) -- (m2);
    \end{tikzpicture}
    \end{center}
  \end{description}
\end{definition}

\noindent In the following, we use the function $\spo$ (of Definition~\ref{def:dpo}) to chain transitions by span composition (see diagrams below). Given two spans $\vectorf{f}=\vectorm{f_1}{f_2}$ and $\vectorf{g}=\vectorm{g_1}{g_2}$, we define their composition as $\vectorf{g}\vectorf{f} = \vectorm{f_1h_1}{g_2h_2}$ where $\vectorf{h}$ is the pullback of $\vectorm{f_2}{g_1}$. A partial morphism $f:M_1 \pmorph M_3$ is a total morphism from the subgraph $\text{dom}(f)$ of $M_1$ to $M_3$, that is $f:M_1 \supseteq \text{dom}(f) \to M_3$. Given a span $M_1\overset{l}{\leftarrow} D \overset{r}{\rightarrow} M_3$, its corresponding partial morphism, denoted $M_1\pmorph M_3$, is defined on $l(D)$ as $l^{-1}r$ and undefined otherwise~\cite{Ehrig:SPO}.

\begin{center}
  \begin{tikzpicture}[scale=0.8]
    \node (g1) at (-2,0) {\(M_1\)};
    \node (g2) at (0,0) {\(D_1\)};
    \node (g3) at (2,0) {\(M_2\)};
    \node (g4) at (4,0) {\(D_2\)};
    \node (g5) at (6,0) {\(M_3\)};
    \node (g) at (2,-1) {\(D\)};
    \draw [->] (g2) -- node [above] {\(f_1\)} (g1);
    \draw [->] (g2) -- node [above] {\(f_2\)} (g3);
    \draw [->] (g4) -- node [above] {\(g_1\)} (g3);
    \draw [->] (g4) -- node [above] {\(g_2\)} (g5);
    \draw [->] (g) -- node [below,near end] {\(h_1\)} (g2);
    \draw [->] (g) -- node [below,near end] {\(h_2\)} (g4);
    \end{tikzpicture}
  \qquad
    \begin{tikzpicture}[scale=0.8]
    \node (g1) at (0.5,0) {\(M_1\)};
    \node (g5) at (3.5,0) {\(M_3\)};
    \node (g) at (2,0) {\(D\)};
    \draw [->] (g) -- node [above] {\(l\)} (g1);
    \draw [->] (g) -- node [above] {\(r\)} (g5);
    \draw [->] (g1) to [bend right] (g5);
  \end{tikzpicture}
\end{center}

\begin{definition}[Causality]
  \label{def:causality}
  Let $t_1:M_1\overset{m_1,\vectorf{r_1}}{\Rightarrow} N_1$ and $t_2:M_2\overset{m_2,\vectorf{r_2}}{\Rightarrow} N_2$ be two transitions bracketing a trace $\theta:t_1;t_1';t_2';\cdots ;t_n';t_2$. The rules inducing $t_i, i\in\{1,2\}$, are $\vectorf{r_i}=L_i{\remb} K_i {\lemb} R_i$ with matchings $m_i\in\homg$ into $M_1$ and $M_2$, respectively.
  \begin{description}
  \item[enablement]
    \label{def:dep}
    Let $\vectorf{g}$ be a span such that $\vectorf{r_1}\posI{g}\vectorf{r_2}$. If the diagram below on the left commutes then $t_1$ enables $t_2$, denoted $t_1 <_{\theta} t_2$.
    In the diagram below on the left, the partial morphism $N_1\pmorph M_2$ is obtained from the composition of $\spo(t_1')\circ\dots\circ\spo(t_n')$.
  \item[prevention]
    \label{def:inhibition}
    Let $\vectorf{g}$ be a span such that $\vectorf{r_2}\negI{g} \vectorf{r_1}$. If the diagram below on the right commutes then $t_2$ prevents $t_1$, denoted $t_2 \dashv_{\theta} t_1$.
    In the diagram below on the right the partial morphism $M_1\pmorph M_2$ is the composition of $\spo(t_1)\circ\spo(t_1')\circ\dots\circ\spo(t_n')$.
  \end{description}
\end{definition}

\begin{center}
  \begin{tikzpicture} [scale=0.7]
    \node (o) at (1.5,0) {\(O\)};
    \node (m1) at (-3,3) {\(M_1\)};
    \node (d1) at (-1.5,3) {\(D_1\)};
    \node (n1) at (0,3) {\(N_1\)};
    \node (n2) at (6,3) {\(N_2\)};
    \node (d2) at (4.5,3) {\(D_2\)};
    \node (m2) at (3,3) {\(M_2\)};
    \node (r1) at (0,1) {\(R_1\)};
    \node (k1) at (-1.5,1) {\(K_1\)};
    \node (l1) at (-3,1) {\(L_1\)};
    \node (l2) at (3,1) {\(L_2\)};
    \node (k2) at (4.5,1) {\(K_2\)};
    \node (r2) at (6,1) {\(R_2\)};
    \node (t) at (1,-1) {\small{enablement}};
    \draw [->] (o) -- node [below] {\small{\(g_1\)}} (r1);
    \draw [->] (o) -- node [below,midway] {\small{\(g_2\)}} (l2);
    \draw [->] (r1) --  (n1);
    \draw [->] (l2) --  (m2);
    \draw [->] (n1) --  (m2);
    \draw [->] (l1) --  (m1);
    \draw [->] (r2) --  (n2);
    \draw [->] (k1) -- (d1);
    \draw [->] (k2) -- (d2);
    \draw [->] (k1) -- (r1);
    \draw [->] (k1) -- (l1);
    \draw [->] (k2) -- (l2);
    \draw [->] (k2) -- (r2);
    \draw [->] (d1) -- (m1);
    \draw [->] (d1) -- (n1);
    \draw [->] (d2) -- (m2);
    \draw [->] (d2) -- (n2);
  \end{tikzpicture}
  \hspace*{0.6cm}
  \begin{tikzpicture}[scale=0.7]
    \node (o) at (0,0) {\(O\)};
    \node (m1) at (-2,3) {\(M_1\)};
    \node (m2) at (2,3) {\(M_2\)};
    \node (l1) at (-2,1) {\(L_1\)};
    \node (l2) at (2,1) {\(L_2\)};
    \node (t) at (0,-1) {\small{prevention}};
    \draw [->] (o) -- node [below] {\footnotesize{\(g_1\)}} (l1);
    \draw [->] (o) -- node [below,near end] {\footnotesize{\(g_2\)}} (l2);
    \draw [->] (l1) --  (m1);
    \draw [->] (l2) --  (m2);
    \draw [->] (m1) --  (m2);
  \end{tikzpicture}
 \end{center}
To make the underlying span explicit, we sometimes write $(t_1,t_2,\vectorf{g})\in <_{\theta}$ and $(t_1, t_2, \vectorf{g})\in \dashv_{\theta}$.
  \begin{center}
    \includegraphics[scale=0.2]{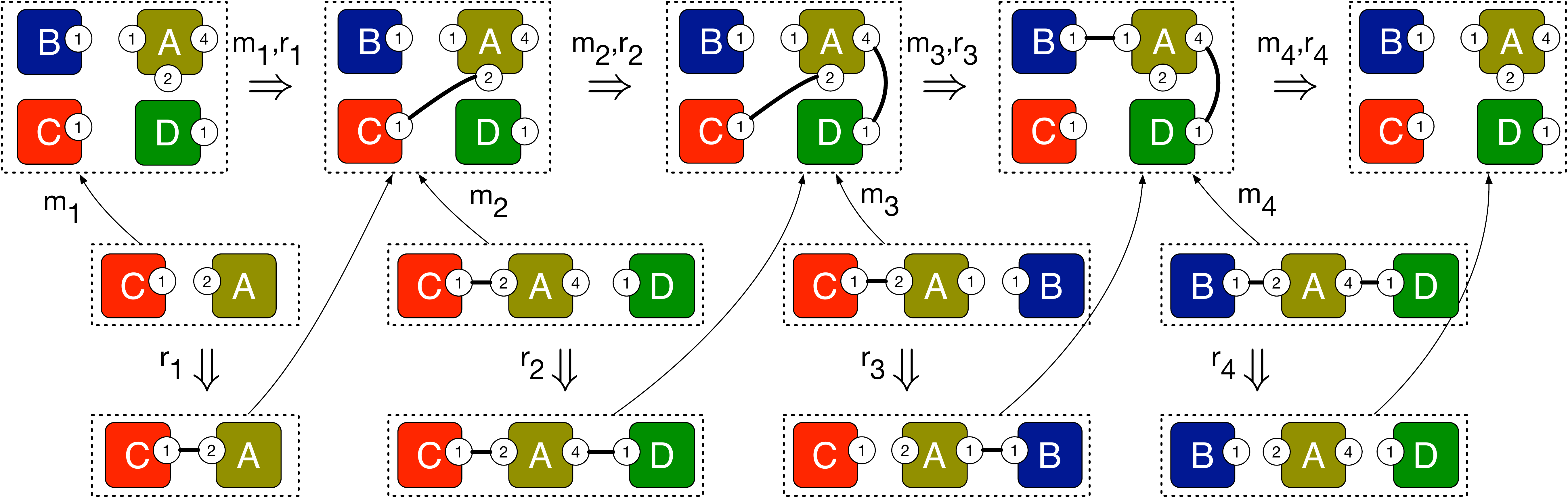}
    \captionof{figure}{Transition $t_2$ binds agents $\no{A}$ and $\no{D}$, needed by transition $t_4$. Transition $t_3$ needs to happens between the two, as it (i) binds agents $\no{A}$ and $\no{B}$ as needed by $t_4$ and (ii) unbind $\no{A}$ from $\no{C}$ which was necessary for $t_2$.}
    \label{fig:trace}
  \end{center}

\begin{example}
  \label{ex:low_res}
Consider the trace of Figure~\ref{fig:trace}. The first transition enables the second and third transitions. The second transition enables the last transition. However, it is a ``delayed" enabler: it partially fulfills the precondition of the last transition, but the third transition has to happen before. Note that such type of causality is not captured by the graph rewriting framework of~\cite{AlgebraicGR,BaldanThesis,Lambers2006}\footnote{For immediate transitions, enablement and prevention coincide with the sequential dependence and critical pairs, respectively, of Refs.~\cite{AlgebraicGR,BaldanThesis,Lambers2006}. See the appendix for more details.}. Lastly note that the third transition is a preventer for the second one.
\end{example}

\section{Posets of graph rewriting events}
\label{sec:posets_of_graphs}

In this section we abstract a trace into a poset of events, and concretize a poset back into a set of traces. Each transition becomes an event in a poset with the underlying rule as its label. Similarly, in the concretization, each event in a poset corresponds to a transition such that transitions compose into a trace. The abstraction is used to reduce the number of simulation traces to a small set of posets, and the concretization recomputes a ``representative'' trace from each poset. Concretized traces are used in the next section.

\subsection{From traces to posets}
\label{sec:concret}

A transition $t$ (Definition~\ref{def:dpo}) is a pair of spans---$\spo(t)=M\remb D\lemb N$ and the underlying rule $\vectorf{r}: L\remb K\lemb R$---and a matching $m:L\to M$. When abstracting a trace into a partial order, we drop the span $\spo(t)$ and $m$. The enabling and prevention relations between transitions in a trace (Section~\ref{sec:ts}) translate into a partial order on events, labeled by the underlying rules.

We proceed in two steps. Enabling and prevention between transitions hinge on positive and negative influence between the underlying rules (see Definition~\ref{def:causality}). Recall that when transition $t$ enables transition $t'$ within a trace $\theta$ there exists a span $\vectorf{f}$ for which $(t, t',\vectorf{f})\in <_{\theta}$.
The first abstraction, $\absa$, forgets the matching and the span $\spo(t)$ of a transition $t$, but preserves enablement and prevention relations between transitions and the positive and negative influence between the underlying rules.

\begin{definition}[Abstraction step 1]
  \label{def:abs1}
   Let $\theta=t_1;t_2;\cdots ;t_n$ be a trace and $E=\{e_1,e_2,\cdots,e_n\}$\footnote{We can define a function $\text{id}:T\to \nat$ from transitions to natural numbers such that $\text{id}(t_i) = i$. The set of events is then $E=\{1,2,\cdots n\}$.} be a set of events. Events are labeled using a function $\labl:E\to\mathcal{R}$ such that $\labl(e_i) = r_i$ if $t_i: M_i\overset{m_i,\vectorf{r_i}}{\Rightarrow} N_i$, for $i\leq n$.
We then define two relations $\redld{+\cdot},\redld{-\cdot}\subseteq E\times E\times\spang$:
\begin{center}
  $e_i\posI{f} e_j \iff (t_i, t_j,\vectorf{f})\in <_{\theta}$ and
  $e_i\negI{f} e_j \iff (t_i, t_j,\vectorf{f})\in \dashv_{\theta}$,
\end{center}
for $e_i,e_j\in E$, $i,j\leq n$ and $\vectorf{f}\in\spang$. We denote this first abstraction of $\theta$ with $\absa(\theta)=(E,\labl,\redl{+},\redl{-})$.
\end{definition}

The notation $e\posI{f}e'$, for some span $\vectorf{f}$, overloads the notation $\labl(e)\posI{f}\labl(e')$. Keep in mind, however, that the first is \emph{defined} on events whereas the second can be \emph{inferred} from the rules on which it holds (see Definition~\ref{def:pos_infl}).

In the second abstraction step, $\absb$, we map the relations $\redl{+}$ and $\redl{-}$ to corresponding partial orders on events. This step simply forgets the spans responsible for the enablement and prevention relations on transitions.

\begin{definition}[Abstraction step 2]
  \label{def:abs2}
  Let $E$ be a set of events equipped with a labeling function $\labl:E\to\mathcal{R}$ and two relations $\redld{+\cdot},\redld{-\cdot}\subseteq E\times E\times\spang$.
We translate the relations on events from Definition \ref{def:abs1} into two new relations $<,\Vdash \subseteq E\times E$:
\vspace{-0.5cm}
\[
e_i< e_j\iff e_i\posI{f} e_j\text{ and }
e_i \Vdash e_j \iff e_j\negI{f} e_i.
\]
The associated poset is defined as $\absb(E,\labl,\redl{+},\redl{-}) = (E,\labl,\leq,\vdash)$, where $\leq$ and $\vdash$ are the transitive and reflexive closure of $<$ and $\Vdash$, respectively. We call the two relations $\leq$ and $\vdash$, \emph{(enabling) precedence} and \emph{non-enabling precedence}, respectively\footnote{In order to not introduce unnecessary terminology, we abuse the term \emph{poset} to mean the structure $(E,\labl,\leq,\vdash)$ where the set of events $E$ is equipped with \emph{two} partial orders. We could instead define $(E,\labl,(<\cup\Vdash)^{\star})$ but in this case we forget the distinction between $<$ and $\Vdash$.}.
\end{definition}

\begin{lemma}
  \label{lem:abs2_order}
  Let $\theta$ be a trace and let $e,e'\in E$ be two events with $\absb\absa(\theta)=(E,\labl,\leq,\vdash)$. If $e< e'$ then there exists a span $\vectorf{f}$ such that $\labl(e)\posI{f}\labl(e')$. Similarly, if $e\Vdash e'$, then there exists a span $\vectorf{f}\in\spang$ such that $\labl(e')\negI{f}\labl(e)$.
\end{lemma}

A morphism on posets is a function on events that preserves labels and the two precedence relations. An isomorphism between two posets $s_1$ and $s_2$ is denoted by $s_1\iso s_2$. For a set of traces $\Theta = \{\theta_1,\cdots,\theta_n\}$, we write $\mathcal{S}=(s_1,\cdots, s_k)/_{\iso}\text{ with }k\leq n$ for the set of posets obtained via $\absb\absa$ and quotiented by iso.

\begin{example}
\label{example:abstract_trace}
Consider the trace $\theta=t_1;t_2;t_3;t_4$ of Example~\ref{ex:low_res}. The corresponding poset consists of the events $\{e_1,e_2,e_3,e_4\}$ with the relations
$< = \{(e_1,e_2); (e_1,e_3);$ $(e_2, e_4); (e_3,e_4)\}$ and $\Vdash =\{(e_2,e_3)\}$. Note that $e_2$ is a non-enabling precedent of $e_3$, as in the original trace transition $t_3$ prevents transition $t_2$. 
\end{example}

\subsection{From posets to traces}
\label{sec:refinement}
We next specify the concretization from posets to traces. Again, we proceed in two steps. The first concretization retrieves the intermediate structure $(E,\labl,\redl{+},\redl{-})$ from a poset $(E,\labl,\leq,\vdash)$. This step recovers the influence (positive or negative) between the rules underlying two events that are in a particular precedence or non-enabling precedence relation.

\begin{definition}[Concretization step 1]
  \label{def:concr1}
  Let $(E,\labl,\leq,\vdash)$ be a poset. We define the relations $\redl{+},\redl{-}\subseteq E\times E\times\spang$ as follows:
  \begin{itemize}
  \item $e_i\posI{f} e_j \iff \labl(e_i)\posI{f}\labl(e_j)$ and $e_i < e_j$, for some $\vectorf{f}\in\spang$;
  \item $e_i\negI{f} e_j \iff \labl(e_i)\negI{f}\labl(e_j)$ and $e_j \Vdash e_i$, for some $\vectorf{f}\in\spang$
  \end{itemize}
  where $<$ and $\Vdash$ are the reduced relation of $\leq$ and $\vdash$, respectively. The concretization of a poset is then $\concra(E,\labl,\leq,\vdash) = (E,\labl,\redl{+},\redl{-})$.
\end{definition}

\begin{example}
\label{ex:valid_dec}
Consider a poset of events $e_1,e_2$ and $e_3$ with labels $\vectorf{r_1}$, $\vectorf{r_2}$ and $\vectorf{r_3}$,
respectively, as shown in Figure~\ref{fig:posets25}. Furthermore, suppose that events $e_1$ and $e_2$ both precede $e_3$. For the pair $e_1< e_3$, one can infer the positive influence $\vectorf{r_1}\posI{f} \vectorf{r_3}$. For the pair $e_2< e_3$, we need to consider two possibilities: either
\begin{center}
    \includegraphics[scale=0.2]{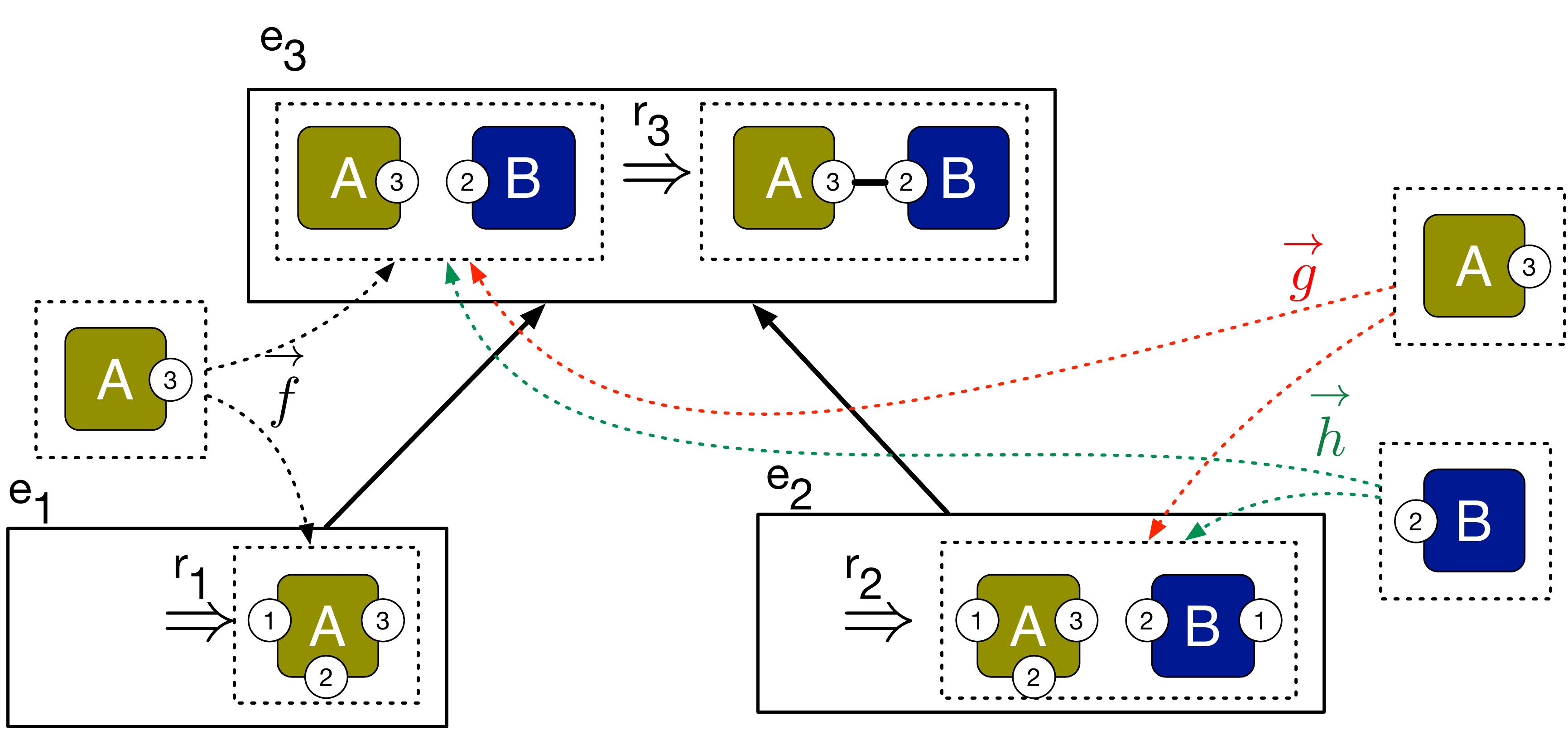}
  \captionof{figure}{All possible influences between three rules.}
  \label{fig:posets25}
\end{center}
$\vectorf{r_2}\posI{g}\vectorf{r_3}$ or $\vectorf{r_2}\posI{h}\vectorf{r_3}$. The relation
induced by the span $\vectorf{g}$ is problematic.
Intuitively, the events $e_1$ and $e_2$ should produce a distinct set of agents for event $e_3$. Specifically, both cannot produce the \emph{same} agent $\no{A}$ that binds to $\no{B}$ in $e_3$. The \emph{consistent} span attributes the creation of agent $\no{A}$ to $e_1$ and the creation of $\no{B}$ to $e_2$ (in addition to a further $\no{A}$ not used in $e_3$). In this manner, both $e_1$ and $e_2$ are necessary for the occurence of $e_3$.
\end{example}

As the example indicates, it is not trivial to retrieve the influence between events from the influence between rules. The problem is that influence between events is a \emph{global} property of the poset, whereas influence between rules is \emph{local} to the two rules. Lack of space prevents us from characterizing the correct concretizations of a poset. Informally, a concretization of a poset $s$ is correct if (i) every relation on events in $s$ is due to a shared resource (i.e. an agent or an edge) and if (ii) every resource in $s$ is consistent throughout $s$.

The second concretization maps events into transitions such that: (i) the transitions compose into a valid trace and (ii) the relations defined on the events hold on the transitions of the trace. We call a \emph{candidate for concretization} any function from events to transitions that satisfies condition (i).

\begin{definition}
  Let $E$ be a set of events with a labeling function $\labl:E\to\mathcal{R}$ and a total order on events $\sqsubset \subseteq E\times E$. A function $\concre:E\to T$ is called a \emph{candidate for concretization} if $\concre(e) = M\overset{m,\vectorf{r}}{\Rightarrow}N$ such that $\labl(e) = \vectorf{r}$, for some graphs $M, N$, and a morphism $m$. Moreover $\concre(e_1);\cdots;\concre(e_n)$, with $e_i\sqsubset e_{i+1}$, $i\leq n$, compose into a trace.
\end{definition}
Any such function must also satisfy condition (ii), as in the following definition.
\begin{definition}[Concretization step 2]
  \label{def:concr2}
  Let $E$ be a set of events equipped with a function $\labl:E\to\mathcal{R}$ and two relations $\redld{+\cdot},\redld{-\cdot}\subseteq E\times E\times\spang$. Let $\sqsubset$ be a total order on events and let $\concre:E\to T$ be a function from events to transitions such that the following hold:
  \begin{itemize}
  \item
    $(t_i, t_j,\vectorf{f})\in <_{\theta} \iff e_i\posI{f} e_j$ and
  \item
    $(t_i, t_j,\vectorf{f})\in \dashv_{\theta} \iff e_i\negI{f} e_j$
  \end{itemize}
  for $e_i,e_j\in E$, $i,j\leq n$. Then the concretized trace is $\concrb(E,\labl,\redl{+},\redl{-},\concre,\sqsubset)$ $=$ $\concre(e_1);\cdots ;$ $\concre(e_n)$, for $e_i\sqsubset e_{i+1}$, $i\leq n$.
 \end{definition}

For $(E,\labl,\leq,\dashv)$ a poset, we write $\concr(E,\labl,\leq,\dashv)$ for the set of all possible concretisations, i.e. the set of all traces $\theta$ as specified by $\concra(E,\labl,\leq,\dashv)$ and $\concrb(E,\labl,\redl{+},\redl{-},\concre,\sqsubset)$.
We write $(\theta,\concre)\in\concr(E,\labl,\leq,\dashv)$ for the concretization function used in reconstructing a particular $\theta$.

\begin{theorem}
  \label{th:concretisation}
  Let $\theta$ be a trace. Then $\theta\in\concr\absb\absa(\theta)$. Moreover, for any trace $\theta'\in\concr\absb\absa(\theta)$, $\absb\absa(\theta)\iso\absb\absa(\theta')$.
\end{theorem}


\vspace*{-0.2cm}
\section{A logic on posets}
\label{sec:posets}

\begin{wrapfigure}{R}{0.55\textwidth}
\centering
\vspace*{-30pt}
{\small
\begin{align*}
  x ::= & x^e ~|~ x^s &\tag{variables on events and posets} \\
  t^s ::= & x^s ~|~ s &\tag{terms on posets} \\
  t^e ::= & x^e ~|~ e & \tag{terms on events}\\
  t ::= & t^s ~|~ t^e & \tag{terms}\\
  \varphi ::= & \exists x.\varphi(x) ~|~ \forall x.\varphi(x) ~|&\tag{quantifiers}\\
  & \neg \varphi ~|~\varphi_1 \wedge \varphi_2~|&\tag{logical connectors}\\
  & t^e\in t^s ~|~ \labl(t^e) = \vectorf{r} ~|~ t^e_1 \leq_{t^s} t^e_2 ~|~ t^e_1 \vdash_{t^s} t^e_2\\
  &~|~ t^e_1\in t^s_1 \arrpos{-} t^e_2\in t^s_2~|~ t^e_1\in t^s_1 \arrpos{+} t^e_2\in t^s_2 & &\tag{predicates}
\end{align*}}
\caption{\label{fig:grammar} The grammar of the poset logic.}
\end{wrapfigure}
In Figure~\ref{fig:grammar} we define a fragment of a first order logic that can be used to express assertions about positive and negative influence between mechanisms, that is, posets. We interpret the logic on the set of posets $\mathcal{S}$, ranged over by $s$, and on the set of events $\mathcal{E}=\cup_{s_i\in S} E_{s_i}$, where $E_{s_i}$ is the set of events in $s_i$. To distinguish between the partial orders of different posets in $\mathcal{S}$, we write $s=(E_s,\leq_s,\vdash_s,\labl_s)$. In the following, $x$ stands for variables, $t$ for terms and the superscripts $e$ and $s$ indicate whether the variables and terms range over events or posets, respectively. Formulas are denoted by $\varphi$ and are built from predicates on variables and terms.

A \emph{valuation} for $\varphi$ is a function $v:\text{fv}(\varphi)\to\mathcal{E}\uplus\mathcal{S}$ from the set of free variables of $\varphi$ to the set of events and posets. The \emph{evaluation} of $\varphi$ is defined below and requires a valuation function $v$ for the set of free variables of $\varphi$; the evaluation is therefore parametric on $v$. We use two functions, one to evaluate terms $\enct{}_v:t\to\mathcal{E}\uplus\mathcal{S}$ and one to evaluate formulas $\enc{}_v:\varphi\to\set{T,F}$. A formula $\varphi$ is satisfiable if there exists $v$ such that $\enc{\varphi}_v$ evaluates to true. The interpretation of formulas and terms is shown in Figure~\ref{fig:interpret}.

\begin{figure}[H]
\centering
\vspace*{-25pt}
{\small
\begin{align*}
  \enc{\forall x^s.\varphi}_{v} & \iff\text{ for all }s\in\mathcal{S}, \enc{\varphi(s/x)}_{v}\\
  \enc{\exists x^s.\varphi}_{v} & \iff\text{ for some }s\in\mathcal{S}, \enc{\varphi(s/x)}_{v} \\
  \enc{\neg\varphi}_v &= \neg\enc{\varphi}_v \\
  \enc{\varphi_1\wedge\varphi_2}_v &= \enc{\varphi_1}_v\wedge\enc{\varphi_2}_v\\
  \enc{t^e\in t^s}_v &\iff\enct{t^e}_v\in\enct{t^s}_v\\
  \enc{\labl(t^e) = \vectorf{r}}_v &\iff\labl(\enct{t^e}_v)=\vectorf{r}\\
  \enc{t^e_1 \leq_{t^s} t^e_2}_v &\iff e_1 \leq_s e_2\text{ where }e_1 = \enct{t^e_1}_v,e_2 = \enct{t^e_2}_v,s = \enct{t^s}_v\\
  \enc{t^e_1 \vdash_{t^s} t^e_2}_v &\iff e_1 \vdash_s e_2\text{ where }e_1 = \enct{t^e_1}_v,e_2 = \enct{t^e_2}_v,s = \enct{t^s}_v\\
  \enc{t^e_1\in t^s_1 \arrpos{+/-} t^e_2\in t^s_2}_v &\iff e_1\in s_1 \arrpos{+/-} e_2\in s_2
  \text{ where }
  e_1 = \enct{t^e_1}_v,e_2 = \enct{t^e_2}_v,\\
  &\hspace{5cm} s_1 = \enct{t^s_1}_v,s_2 = \enct{t^s_2}_v\\
  \enct{x}_{v} &= v(x)\\
  \enct{e}_{v} &= e\\
  \enct{s}_{v} &= s
\end{align*}
}%
\caption{\label{fig:interpret} The intepretation of the poset logic.}
\end{figure}


\begin{example}
We return to the introductory example. The mechanisms of binding an agent $\no{A}$ to an agent $\no{X}$ or to an agent $\no{Y}$ consist in the application of rule $\vectorf{r_{\text{AX}}}$ and $\vectorf{r_{\text{AY}}}$, respectively. The assertion that the first mechanism prevents (or conflicts with) the second is written as $\exists e_1.(e_1\in s_1\wedge\labl(e_1)=\vectorf{r_{\text{AX}}})\wedge\exists e_2.(e_2\in s_2\wedge\labl(e_2)=\vectorf{r_{\text{AY}}})\wedge e_1\in s_1\arrpos{-}e_2\in s_2$. The logic allows us to formulate more complex mechanisms. For our example, we can write $\exists e.e\in s\wedge\labl(e)=\vectorf{r_{\text{AX}}}\vee\labl(e)=\vectorf{r_{\text{AY}}}$ for a mechanism that produces $\no{A}$ bound to either $\no{X}$ or $\no{Y}$.
\end{example}
%
The predicates $e_1\in s_1\arrpos{+} e_2\in s_2$ and $e_1\in s_1\arrpos{-} e_2\in s_2$ check for enablement and prevention between two posets. Informally, $e_1$ and $e_2$ represent the "meeting point" of the two posets $s_1$ and $s_2$.
We use these events to reconstruct a graph that represents a context in which $s_1$ enables or prevents $s_2$.

The \emph{causal past} of an event is the set of events that preceded it. We denote with $[e]_s$ the causal past of an event $e\in E_s$ and define $[e]_s=(E',\leq',\vdash',\labl')$ with $E'=\{e': e'\in E,e'\leq e\}$ and $\leq',\vdash',\labl'$ defined like $\leq,\vdash,\labl$ but restricted to $E'$.

\begin{definition}[Occurrence context of an event in a poset]
Let $s$ be a poset and let $e\in E_s$ be an event.
Furthermore, let $(\theta,\concre)\in\concr([e]_s)$ be a concretization of $[e]_s$.
We say that a morphism $m$ is an \emph{occurrence context of $e$ in $s$} if $\concre(e) = M\overset{m,\labl(e)}{\Rightarrow}N$, for some graphs $M,N$.
\end{definition}

\begin{wrapfigure}{R}{0.25\textwidth}
\centering
\vspace*{-25pt}
  \begin{tikzpicture} [scale=0.8]
    \node (o) at (0,-1) {\(O\)};
    \node (n) at (0,2) {\(M\)};
    \node (l1) at (-1.3,-0.5) {\(L_1\)};
    \node (l2) at (1.3,-0.5) {\(L_2\)};
    \node (n1) at (-1,1) {\(M_1\)};
    \node (n2) at (1,1) {\(M_2\)};
    \draw [->] (l1) -- node [left] {\(m_1\)} (n1);
    \draw [->] (l2) -- node [right] {\(m_2\)} (n2);
    \draw [->] (o) -- node [below,midway] {\(f_1\)} (l1);
    \draw [->] (o) -- node [below,midway] {\(f_2\)} (l2);
    \draw [->] (n1) -- (n);
    \draw [->] (n2) -- (n);
    \draw [->] (o) to [bend left] node [right] {\(g_1\)} (n1);
    \draw [->] (o) to [bend right] node [left] {\(g_2\)} (n2);
  \end{tikzpicture}
\end{wrapfigure}
The occurrence context of $e_1$ in $s_1$ and of $e_2$ in $s_2$ is specified by matchings $m_1:L_1\to M_1$ and $m_2:L_2\to M_2$, respectively. The diagram on the right illustrates the prevention of $s_2$ by $s_1$.  Since the graph $M$ contains both $M_1$ and $M_2$, both events $e_1$ and $e_2$ can occur in that context. We then say that $M$ is a \emph{scenario} for the prevention of $s_2$ by $s_1$, which is induced by a negative influence between the underlying rules, $\labl(e_1)\negI{f}\labl(e_2)$. The scenario graph $M$ is formally defined as follows.

\begin{definition}[Scenario for prevention]
\label{def:scenario}
Let $m_i$ be an occurrence context of event $e_i$ in the poset $s_i$, $i\in\{1,2\}$. Let $\vectorf{f}$ be a span such that $\labl(e_1)\negI{f}\labl(e_2)$.
Define the span $\vectorf{g}= \vectorm{g_1}{g_2}$ as $g_i=m_if_i$, $i\in\{1,2\}$.
We say that the graph $M$ obtained by the pushout $\vectorf{g}$ is a \emph{scenario} (graph) for the prevention of $e_2\in s_2$ by $e_1\in s_1$.
\end{definition}
\begin{example}
  Let $L_1$, $L_2$ be the left hand sides of rules $r_{\text{AX}}$ and $r_{\text{AY}}$ from Figure~\ref{fig:conflict}. We have a negative influence between the rules $r_{\text{AX}}$ and $r_{\text{AY}}$ induced by the agent $\no{A}$.
  The occurence context of $e_{\text{AX}}$ in the poset $\text{AX}$ is obtained from the concretization of the poset $\text{AX}$ and consists of the morphism $L_1\to G_{\text{AX}}$. Similarly the occurence context of $e_{\text{AY}}$ in the poset $\text{AY}$ is $L_2\to G_{\text{AY}}$.
  There is no scenario for prevention as the graph $G$ (in Figure~\ref{fig:conflict}) is not a site-graph.
\end{example}

In a similar manner we interpret the enabling relation between two mechanisms.
The predicate $(e_1\in s_1\arrpos{+/-} e_2\in s_2)$ returns true if there exists a scenario $M$ as defined above. The pushout does not always exists and, in consequence, mechanisms do not always interact with one another.

The logic is implemented as a systematic inspection of each poset. The set of posets does not have in itself a structure, and therefore there is no smart strategy for deciding whether a formula holds. The point of the logic is to give a formal language and an interpretation for influence between posets.

\begin{example}

  Let us look at a Kappa model slightly more complicated than the one in the Introduction. We give the rules in the figure below. The two posets build up the graphs $G_{\text{AX}}$ and $G_{\text{AY}}$. Then there are two ``resources'' which can produce an inhibition between the two posets. They produce two scenario graphs for inhibition $G_1$ and $G_2$, shown in Figure~\ref{fig:another}.
  \begin{figure}
  \begin{center}
    \includegraphics[scale=0.2]{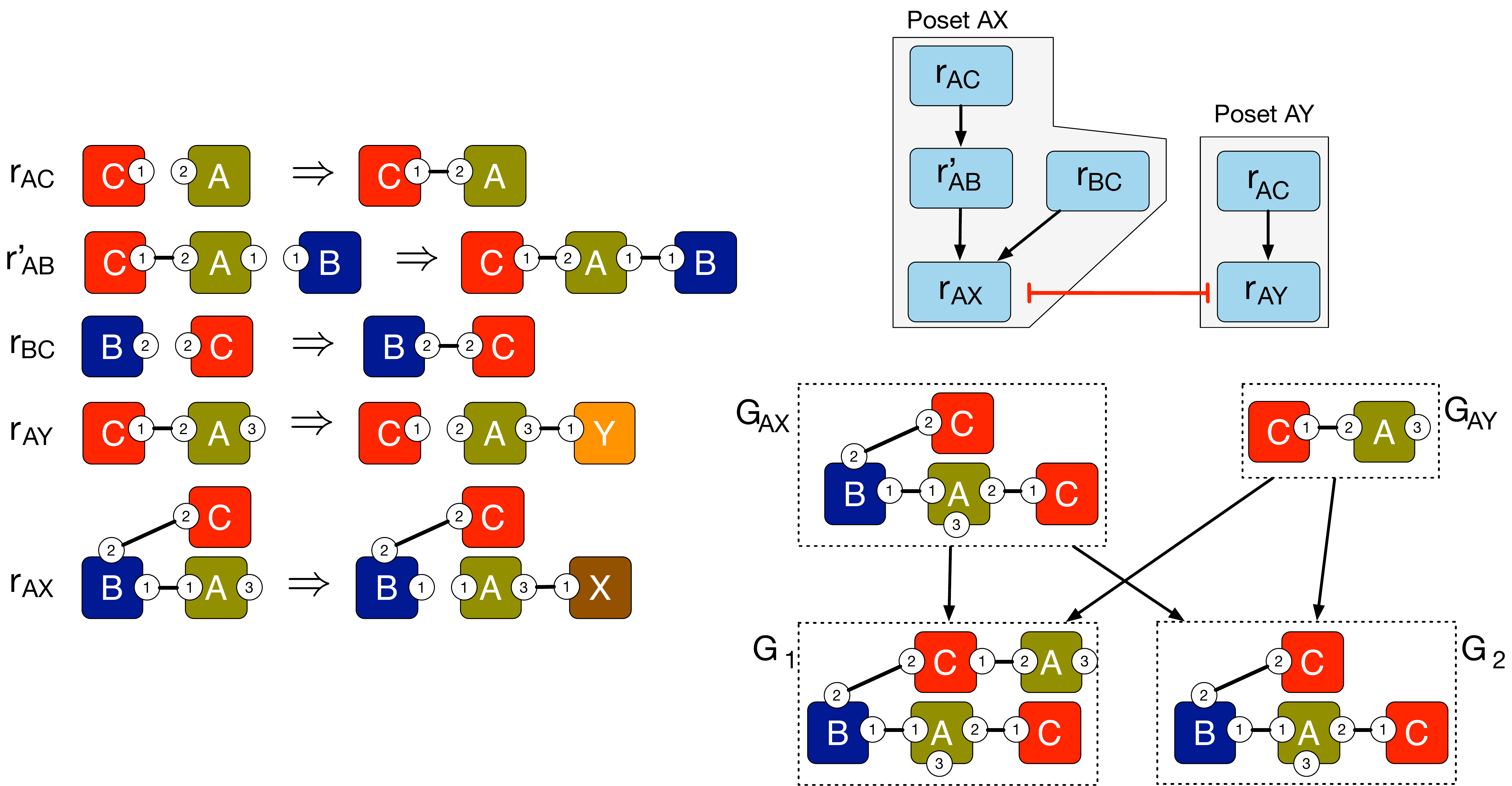}
  \end{center}
  \caption{\label{fig:another} A Kappa model for which there are two scenarios for the prevention between the events labeled $r_{\text{AX}}$ and $r_{\text{AY}}$.}
  \end{figure}

Let us change rule $r'_{\text{AB}}$ into $r''_{\text{AB}}$ and keep everything else the same. With the new rule the graph build up by Poset AX requires the site $2$ of agent $C$ to be free. In this case only one scenario for inhibition can still occur, shown in Figure~\ref{fig:yetanother}.

\begin{figure}
  \begin{center}
  \includegraphics[scale=0.2]{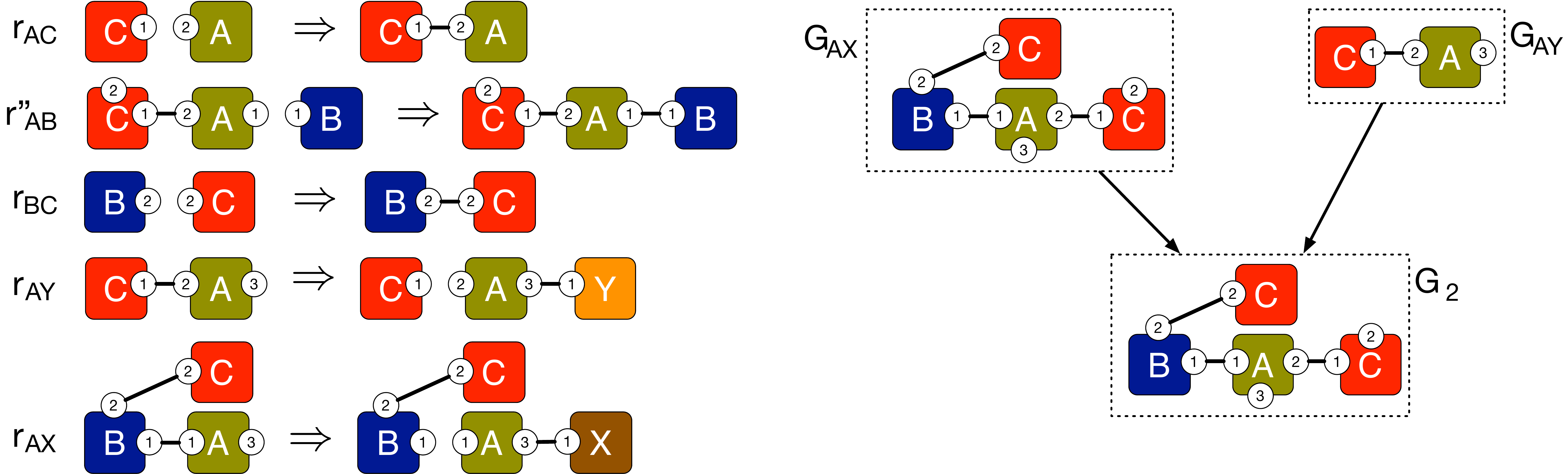}
  \end{center}
   \caption{\label{fig:yetanother} A slightly different Kappa model for which only one of the scenarios is still valid.}
\end{figure}
\end{example}

\section{Conclusions}
\label{sec:concl}

Given a categorical notion of graph rewrite system, we defined positive and negative influence between rules. This allowed us to define sequential and parallel independence between state transitions and the relations of enablement and prevention. These were then lifted to the poset abstraction of a trace of state transitions, where they became enabling and non-enabling precedence relations \emph{within} a poset. The formulation of a logic on posets then allows us to formulate questions about enablement and prevention relations \emph{between} posets. We ended by specifying how the concretization of posets back into a trace provides a scenario graph that establishes the truth (or falsity) of a statement about poset interaction. These notions, together with their implementation, are meant to assist a modeler in checking the consistency between observations and the mechanisms that are implied by a rule-based model.

\emph{Acknowledgements.}
We gratefully acknowledge illuminating discussions with Russ Harmer, Jerome Feret, and Jonathan Laurent. Special thanks to Pierre Boutillier for his help in developing and integrating the model checker resulting from this contribution into the Kappa software framework.


\bibliographystyle{eptcs}
\bibliography{bio.bib}


\end{document}